\documentclass[aps,prd,
preprintnumbers,groupedaddress]{revtex4}
\usepackage{epsbox}
\usepackage{amsmath}
\usepackage[dvips]{graphicx}

\begin{document}
\preprint{\vtop{
{\hbox{YITP-07-22}\vskip-0pt
                 \hbox{KANAZAWA-07-02} \vskip-0pt
}
}
}

\author{
Kunihiko Terasaki   
}

\title{ 
${\bf D_{s0}^+(2317)}$ as an Iso-triplet Four-quark Meson and \\
Production of Its Neutral and Doubly Charged Partners\footnote
{Invited talk at the workshop, YKIS2006, on "New Frontiers on QCD" 
-- Exotic Hadrons and Hadronic Matter --, November 20 - December 8, 
2006, at the Yukawa Institute for Theoretical Physics, Kyoto 
University, Kyoto, Japan. 
} 
}

\affiliation{
Yukawa Institute for Theoretical Physics, Kyoto University,
Kyoto 606-8502, Japan\\
Institute for Theoretical Physics, Kanazawa University, 
Kanazawa 920-1192, Japan
}

\begin{abstract}{
By studying $D_{s0}^+(2317)\rightarrow D_s^+\pi^0$ and 
$D_{s0}^+(2317)\rightarrow D_s^{*+}\gamma$ decays, it is shown that 
assigning $D_{s0}^+(2317)$ to the iso-triplet four-quark meson 
$\hat F_I^+$ is favored. Productions of its partners $\hat F_I^0$ and 
$\hat F_I^{++}$ are also studied. As the result, it is concluded that 
they could be observed in $B_d^0\rightarrow (D_s^+\pi^-)\bar D^0$ and 
$B_u^+\rightarrow (D_s^+\pi^+)D^-$. Their iso-singlet partner 
$\hat F_0^+$ might have been observed in the radiative 
$B_{u(d)}^{+(0)}\rightarrow \bar D^{0(-)}D_s^{*+}\gamma$ decays by 
the BELLE collaboration. 
}\end{abstract}

\maketitle

\section{Introduction}

Inclusive $e^+e^-$ annihilation 
experiments~\cite{BABAR-D_{s0},CLEO-D_{s0}} have observed a narrow 
($< 4.6$ MeV~\cite{PDG06}) scalar resonance [denoted by 
$D_{s0}^+(2317)$] in the $D_s^+\pi^0$ channel. 
However, no evidence for it has been observed in the $D_s^{*+}\gamma$ 
channel, so that a severe constraint~\cite{CLEO-D_{s0}},  
\begin{equation}
R(D_{s0}^+(2317))
= \frac
{\Gamma(D_{s0}^+(2317) \rightarrow D_{s}^{*+}\gamma)}
{\Gamma(D_{s0}^+(2317) \rightarrow D_{s}^{+}\pi^0)} < 0.059, 
                                     \label{eq:constraint-D_{s0}}
\end{equation}
has been provided. In addition, we here list the measured 
ratio of decay rates~\cite{PDG06} 
\begin{equation}
R(D_{s}^{*+})^{-1}
= \frac{\Gamma(D_{s}^{*+} \rightarrow D_{s}^{+}\pi^0)}
  {\Gamma(D_{s}^{*+} \rightarrow D_{s}^{*+}\gamma)}
=0.062\pm 0.008. 
                                           \label{eq:constraint-D_s^*}
\end{equation}
Eq.~(\ref{eq:constraint-D_s^*}) implies that the isospin non-conserving 
interaction is much weaker than the electromagnetic interaction. 
Therefore, Eq.~(\ref{eq:constraint-D_{s0}}) means that the underlying 
interaction of the decay $D_{s0}^+(2317) \rightarrow D_{s}^{+}\pi^0$ is 
much stronger than the electromagnetic interaction, i.e., it is the 
ordinary strong interaction as is well known. In this case, 
$D_{s0}^+(2317)$ should be an iso-triplet meson which can be realized by 
a four-quark state. 

To confirm the above conjecture, we shortly visit scalar four-quark 
mesons and discuss that charm-strange scalar four-quark mesons can be 
narrow, in {\bf II}, and study their radiative decays and isospin 
non-conserving decays in {\bf III}. Productions of charm-strange scalar 
mesons in $e^+e^-$ annihilation and in hadronic $B$ decays are 
investigated in {\bf IV}. A brief summary is given in the final 
section. 

\section{Charmed Scalar Four-quark Mesons}

Observed low lying scalar mesons~\cite{PDG06}, 
$a_0(980)$, $f_0(980)$, $K_0^*(800)$ and $f_0(600)$, 
can be well understood by the  $[qq][\bar q\bar q]$ states, 
$\hat \delta^s\sim [ns][\bar n\bar s]_{I=1}$,
$\hat \sigma^s\sim [ns][\bar n\bar s]_{I=0}$, 
$\hat \kappa\sim [ud][\bar n\bar s]$,
$\hat \sigma\sim [ud][\bar u\bar d]$, $(n=u,d)$, 
which are dominantly of $\bar 3_c\times 3_c$ of color $SU_c(3)$, as 
suggested long time ago~\cite{Jaffe} and supported at this 
workshop~\cite{supports}. (However, for simplicity, a possible small 
mixing of $6_c\times \bar 6_c$ is ignored in this talk.) 

With this in mind, we replace one of light quarks in 
$[qq][\bar q\bar q]$ by the charm quark $c$. Then we have the 
charmed scalar $[cq][\bar q\bar q]$ mesons, 
$\hat F_I\sim [cn][\bar n\bar s]_{I=1}$,
$\hat F_0^+\sim [cn][\bar n\bar s]_{I=0}$, 
$\hat D^s\sim [cs][\bar n\bar s]$,
$\hat D\sim [cn][\bar u\bar d]$ and 
$\hat E^0\sim [cs][\bar u\bar d]$. 
However, we here study only $\hat F_I^{0,+,++}$ and $\hat F_0^+$. (For 
the other components, see Refs.\cite{Terasaki-D_{s0},TM,Trento}.) 
When we assign~\cite{Terasaki-D_{s0}} $D_{s0}^+(2317)$ to $\hat F_I^+$ 
as conjectured in {\bf I}, one might wonder if it can be so narrow. 
However, its narrow width can be understood by a small rate for the 
dominant decay $\hat F_I^+\rightarrow D_s^+\pi^0$ which is given by a 
small overlap of (color and spin) wavefunctions. Such a small overlap 
can be seen by decomposing a color-singlet scalar four-quark state of 
$\bar 3_c\times 3_c$ into a sum of products of $\{q\bar q\}$ pairs.  
The coefficient of the product of two color- and spin-less 
$\{q\bar q\}$ pairs in the decomposition provides the overlap under 
consideration. Therefore, the parameters describing the overlaps 
between a charm-strange scalar four-quark meson and two pseudoscalar 
mesons, for example, $\hat F_I^+$ (or $\hat F_0^+$) and $D_s^+\pi^0$ 
(or $D_s^+\eta$) is given by $|\beta_0|^2=1/12$, and the corresponding 
one between $\hat F_I^+$ (or $\hat F_0^+$) and $D_s^{*+}\rho^0$ (or 
$\omega$, $\phi$, $\psi$) is provided by $|\beta_1|^2=1/4$. (However, 
in the case of the conventional mesons, the corresponding overlap is 
unity, because their color and spin configuration is unique.) For more 
details, see Refs.~\cite{HT-isospin,Trento,Terasaki-production}. 
To see numerically that $\hat F_I^+$ is narrow, we use a hard pion 
technique in the infinite momentum frame~\cite{suppl}. In this 
approximation, the amplitude for two body decay 
$A({\bf p})\rightarrow B({\bf q})\pi({\bf k})$ 
is given by 
\begin{equation}
M(A\rightarrow B\pi)\simeq \Biggl(
\frac{m_A^2 - m_B^2}{f_\pi}\Biggr)\langle{B|A_{\bar \pi}|A}\rangle, 
                                                  \label{eq:hard-pion}
\end{equation}
where the asymptotic matrix element $\langle{B|A_{\bar \pi}|A}\rangle$ 
has been evaluated in the infinite momentum frame. 
Then, by assigning $a_0(980)$ to $\hat \delta^s$ and using 
$\Gamma(a_0(980)\rightarrow \eta\pi)_{\rm exp}\simeq 60$ MeV  
from the measured peak width~\cite{PDG06} as the input data, a rather 
small rate 
$\Gamma(\hat F_I^+\rightarrow D_s^+\pi^0)_{SU_f(4)}
\simeq 8\,\,{\rm MeV}$  
can be obtained, where the $\eta$-$\eta'$ mixing with the mixing angle 
$\theta_P\simeq -20^\circ$ has been taken. Because the 
spatial wavefunction overlap is in the $SU_f(4)$ symmetry limit at this 
stage, however, it is expected that the amplitude is overestimated by 
about $20 -30$ \%. It can be seen~\cite{Trento} by 
comparing the measured rates for the $D^*\rightarrow D\pi$ decays with 
the estimated ones in which  the measured 
$\Gamma(\rho\rightarrow \pi\pi)_{\rm exp}
= 149.4\pm 1.0$ MeV~\cite{PDG06} 
is adopted as the input data. Taking account for the above symmetry 
breaking, we can get 
$\Gamma(\hat F_I^+\rightarrow D_s^+\pi^0)\sim 3-5$ {MeV}. 
This leads to a sufficiently narrow width of
$\hat F_I^+ = D_{s0}^+(2317)$~\cite{HT-isospin,Trento}. 

\section{
Radiative Decays and Isospin Non-conserving Decays 
}

Since it has been known that the vector meson dominance (VMD) with the 
ideal $\omega$-$\phi$ mixing and the flavor $SU_f(3)$ symmetry for the 
strong vertices works fairly well in the radiative decays of light 
vector mesons~\cite{Terasaki-VMD}, we will extend it to the system 
containing charm quark(s) below. Under the VMD, the amplitude 
$A(V\rightarrow P\gamma)$ can be approximated by  
\begin{equation}
A(V\rightarrow P\gamma)
\simeq
\sum_{V'=\rho^0,\,\omega,\,\phi,\,\psi}
\Biggl[\frac{X_{V'}(0)}{m_{V'}^2}\Biggr]
A(V\rightarrow PV'),           
                                             \label{eq:V-P-gamma-VMD}
\end{equation}
where $X_V(0)$ is the $\gamma V$ coupling strength on the photon 
mass-shell. $X_V$ is dependent on 
photon-momentum-square~\cite{Terasaki-VMD}, and the values of $X_V(0)$ 
have been estimated from the analyses in photoproductions of vector 
mesons on various nuclei~\cite{Leith}. The results are 
$X_\rho(0)=0.033\pm 0.003$ GeV$^2$, 
$X_\omega(0)=0.011\pm 0.001$ GeV$^2$, 
$X_\phi(0)=-0.018\pm 0.004$ GeV$^2$ and
$X_\psi(0)\sim 0.054$ GeV$^2$, 
where the last one has been obtained from 
$d\sigma(\gamma N\rightarrow\psi N)/dt|_{t=0}\simeq 20$ nb/GeV$^2$ 
and
$\sigma_{T}(\psi N) = 3.5 \pm 0.8$ mb~\cite{HLW} 
for the $\psi N$ total cross section. ($N$ denotes a nucleon). 
The $VPV'$ coupling strength can be estimated as 
\begin{equation}
|A(\omega\rightarrow\pi^0\rho^0)|\simeq 18\,{\rm GeV}^{-1},
                                             \label{eq:omega-pi-rho}
\end{equation} 
from the measured rate~\cite{PDG06} 
$\Gamma(\omega\rightarrow \pi^0\gamma)_{\rm exp}=0.757\pm 0.024$ MeV   
by putting $V=\omega$, $P=\pi^0$ and $V'=\rho^0$ in 
Eq.~(\ref{eq:V-P-gamma-VMD}) and by inserting the above $X_{\rho}(0)$ 
into it, because the $\omega\rightarrow\pi^0\gamma$ amplitude is 
dominated by the $\rho^0$ pole. The OZI-rule allowed poles for the 
amplitude $A(D^{*+}\rightarrow D^+\gamma)$ are given by the $\rho^0$, 
$\omega$ and $\psi$ mesons. The relevant $SU_f(4)$ relation  
$-2A(D^{*+}\rightarrow D^+\rho^0)         
=2A(D^{*+}\rightarrow D^+\omega)
= \sqrt{2}A(D^{*+}\rightarrow D^+\psi)=\cdots
=A(\omega\rightarrow \pi^0\rho^0)$  
with Eq.~(\ref{eq:omega-pi-rho}) leads to 
$\Gamma(D^{*+}\rightarrow D^+\gamma)_{SU_f(4)}\simeq 2.4$ keV. 
By comparing the above rate with the measured one~\cite{PDG06} 
$\Gamma(D^{*+}\rightarrow D^+\gamma)_{\rm exp}\simeq 1.5$ keV 
(with $\sim$ 50 \% errors), it is seen~\cite{Trento} that (the VMD with) 
the $SU_f(4)$ symmetry (of spatial wavefunction overlap) again 
overestimates the rate by $\sim 50$ \%, as in {\bf II}. 

Now we study radiative decays of charm-strange mesons. The amplitude 
for $D_s^{*+}\rightarrow D_s^+\gamma$ is dominated by $\phi$ and $\psi$ 
poles. Taking the $SU_f(4)$ symmetry relation,  
$\sqrt{2}A(D_s^{*+}\rightarrow D_s^+\phi)
=\sqrt{2}A(D_s^{*+}\rightarrow D_s^+\psi)=\cdots
=A(\omega\rightarrow \pi^0\rho^0)$, 
and Eq.~(\ref{eq:omega-pi-rho}), we can obtain the rate for the 
$D_s^{*+}\rightarrow D_s^+\gamma$ listed in Table~I. 
\begin{table}[t]
\begin{quote}
Table~I. Radiative decays of charm-strange mesons with the spatial wavefunction 
overlap in the $SU_f(4)$ symmetry. The parameter $\beta_1$ which 
provides the overlap of color and spin wavefunctions is given in 
the text. The input data are taken from Ref.~\cite{PDG06}.  
\end{quote}
\begin{center} 
\begin{tabular}
{l c c l c}

\hline
\hspace{3mm} Decay & \quad Pole(s) \quad
& \quad
$\beta_1$ \quad
&\hspace{10mm} Input Data (keV) & \quad
$\Gamma_{SU_f(4)}$ ({keV})   
\normalsize
\\
\hline
{$D_s^{*+}\rightarrow D_s^+\gamma$} & 
{$\phi,\,\psi$} & \quad {1} \quad
& \quad
{$\Gamma(\omega\rightarrow \pi^0\gamma)_{\rm exp}\,\,=\,757\pm 24$} 
\quad
& \hspace{4mm}
{0.8}\\
\hline
{$\hat F_I^+\,\,\,\rightarrow D_s^{*+}\gamma$} & 
{$\rho^0$} & \quad{$ 
{1}/{4}$} \quad
& \quad 
{$\Gamma(\phi\,\rightarrow a_0\gamma)_{\rm exp}\,\,=0.32\pm 0.03$} 
\quad
& \hspace{0mm}
{45}\vspace{0.2mm}\\
\hline
{$\hat F_0^+\,\,\,\rightarrow D_s^{*+}\gamma$} & 
{$\omega$} & \quad {${1}/{4}$} \quad
& \quad
{$\Gamma(\phi\,\rightarrow a_0\gamma)_{\rm exp}\,\,=0.32\pm 0.03$} 
\quad
& \hspace{4mm}
{4.7}\vspace{0.2mm}\\
\hline
{$D_{s0}^{*+}\rightarrow D_s^{*+}\gamma$} & 
{$\phi,\,\psi$} & \quad{1}\quad
& \quad
{$\Gamma(\chi_{c0}\rightarrow \psi\gamma)_{\rm exp}=\,135\pm 15$} 
\quad
& \hspace{0mm}
{35}\\
\hline
\end{tabular}
\end{center} 
\end{table}
For radiative decays of scalar mesons, we consider typical three cases, 
(i) $S=D_{s0}^{*+}\sim \{c\bar s\}$, (ii) $S=\hat F_0^+$ and (iii) 
$S=\hat F_I^+$. Under the VMD, the amplitude is obtained by replacing 
$(V,\,P)$ in Eq. (\ref{eq:V-P-gamma-VMD}) in terms of $(S,\,V)$.  
In the case (i), the amplitude 
$A(D_{s0}^{*+}\rightarrow D_s^{*+}\gamma)$ is dominated by the $\phi$ 
and $\psi$ poles.   
Using the $SU_f(4)$ relation, 
$2A(D_{s0}^{*+}\rightarrow D_s^{*+}\phi)
=2A(D_{s0}^{*+}\rightarrow D_s^{*+}\psi)=\cdots
=A(\chi_{c0}\rightarrow\psi\psi)$,   
and the input data, 
$\Gamma(\chi_{c0}\rightarrow\psi\gamma)_{\rm exp}=135\pm 15\,$ 
keV~\cite{PDG06}, 
we have the rate for the decay $D_{s0}^{*+}\rightarrow D_s^{*+}\gamma$ 
listed in Table~I. The amplitudes 
$A(\hat F_0^+\rightarrow D_s^{*+}\gamma)$ and 
$A(\hat F_I^+\rightarrow D_s^{*+}\gamma)$ in the cases (ii) and (iii) 
are dominated by the $\omega$ pole and the $\rho^0$ pole, respectively.  
Taking the $SU_f(4)$ relation, 
$A(\hat F_0^{+}\rightarrow D_s^{*+}\omega) 
= A(\hat F_I^{+}\rightarrow D_s^{*+}\rho^0)=\cdots
=A(\phi\rightarrow\hat\delta^{s0}\rho^0)\beta_1$, 
with the overlap parameter $\beta_1$ given in {\bf II} and the
input data, 
$\Gamma(\phi\rightarrow a_0(980)\gamma)_{\rm exp}=0.32 \pm 0.03$ 
keV~\cite{PDG06},  
we have the rates for radiative decays of charm-strange mesons listed 
in Table~I, where the spatial wavefunction overlap is still in the 
$SU_f(4)$ symmetry limit. Then, the ratio of the rate 
${\Gamma(\hat F_I^{+}\rightarrow D_s^{*+}\gamma)_{SU_f(4)}}$ 
in Table~I to 
${\Gamma(\hat F_I^{+}\rightarrow D_s^{+}\pi^0)_{SU_f(4)}}$ estimated 
in {\bf II},  
\begin{equation}
\frac{\Gamma(\hat F_I^{+}\rightarrow D_s^{*+}\gamma)_{SU_f(4)}}
{\Gamma(\hat F_I^{+}\rightarrow D_s^{+}\pi^0)_{SU_f(4)}}
\sim 0.005, 
                                        \label{eq:ratio-F_I}
\end{equation}
satisfies well the constraint Eq.~(\ref{eq:constraint-D_{s0}}). 

Isospin non-conserving decays are now in order. The amplitude for the 
$D_s^{*+}\rightarrow D_s^+\pi^0$ decay can be obtained by putting 
$A=D_s^{*+}$ and $B=D_s^+$ in Eq.~(\ref{eq:hard-pion}). Here we 
assume~\cite{CW} that the isospin non-conservation in decays of 
charm-strange mesons is caused by the $\eta$-$\pi^0$ mixing whose   
mixing parameter $\epsilon$ has been estimated to be~\cite{Dalitz} 
\begin{equation}
\epsilon = 0.0105\pm 0.0013.                      \label{eq:epsilon}
\end{equation}
It is very small and of the order of the fine structure constant 
$\alpha$. This implies that the isospin non-conserving interaction is 
much weaker than the electromagnetic one. The $SU_f(4)$ symmetry of 
asymptotic matrix elements and the $\eta$-$\eta'$ mixing lead to 
$2\langle{D_s^+|A_{\pi^0}|D_s^{*+}}\rangle 
= - \epsilon\sin\Theta\cdot\langle{\pi^+|A_{\pi^+}|\rho^0}\rangle$, 
where $\Theta\simeq 35^\circ$ for the usual $\eta$-$\eta'$ mixing  
angle $\theta_P=-20^\circ$. The size of 
$\langle{\pi^+|A_{\pi^+}|\rho^0}\rangle$ can be estimated to be 
$|\langle{\pi^+|A_{\pi^+}|\rho^0}\rangle|\simeq 1.0$~\cite{suppl} 
from the measured rate~\cite{PDG06}   
$\Gamma(\rho\rightarrow\pi\pi)_{\rm exp}=149.4\pm 1.0$ MeV. 
In this way, we are lead to 
$\Gamma(D_s^{*+}\rightarrow D_s^{+}\pi^0)_{SU_f(4)}
\simeq 0.05\,\,{\rm keV}$.  
Comparing this result with 
$\Gamma(D_s^{*+}\rightarrow D_s^{+}\gamma)_{SU_f(4)}$ in Table~I, 
we obtain 
$R(D_s^{*+})^{-1}\simeq 0.06$.    
This is much smaller than unity, as conjectured in {\bf I}, and 
reproduces well the measurement Eq.~(\ref{eq:constraint-D_s^*}). 
Therefore, the present approach seems to be reliable. 

With this in mind, we consider two cases of the isospin non-conserving 
decays of scalar mesons,   
(i) $S^+ = D_{s0}^{*+}$ 
and 
(ii) $S^+ = \hat F_0^+$.    
The amplitude for the $S^+\rightarrow D_s^+\pi^0$ decay is obtained by 
putting $A=S^+$, $B=D_s^+$ and $\pi=\pi^0$ in Eq.~(\ref{eq:hard-pion}). 
Since this decay is assumed to proceed through the $\eta$-$\pi^0$
mixing as discussed above, we replace the matrix elements, 
$\langle{D_s^+|A_{\pi^0}|D_{s0}^{*+}}\rangle$
and 
$\langle{D_s^+|A_{\pi^0}|\hat F_0^+}\rangle$, 
by the OZI-rule allowed 
$-\epsilon\sin\Theta\cdot\langle{D_s^+|A_{\eta^s}|D_s^{*+}}\rangle$ 
and 
$\epsilon\cos\Theta\cdot\langle{D_s^+|A_{\eta^n}|\hat F_0^+}\rangle$, 
respectively. The $SU_f(4)$ relations of asymptotic matrix elements 
are 
$\langle{D_s^+|A_{\eta^s}|D_s^{*+}}\rangle 
=\langle{K^+|A_{\pi^+}|K_0^{*0}(1430)}\rangle$ in the case (i) and 
$2\langle{D_s^+|A_{\eta^n}|\hat F_0^+}\rangle
= \langle{\pi^+|A_{\eta^s}|\hat \delta^{s+}}\rangle\beta_0$ 
in the case (ii). The size of the former is estimated to be 
$|\langle{K^+|A_{\pi^+}|K_0^{*0}(1430)}\rangle|\simeq 0.29$ 
from the experimental data~\cite{PDG06}, 
$\Gamma(K_0^*(1430)\rightarrow K\pi)_{\rm exp} = 270\pm 24$ MeV, 
and the isospin $SU_I(2)$ symmetry, where it has been assumed that 
$K_0^{*0}(1430)$ is the conventional $^3P_0\,\{d\bar s\}$ 
state~\cite{PDG06}. The latter has already been obtained as
$|\langle{\pi^+|A_{\eta^s}|\hat \delta^{s+}}\rangle|$
$=\sqrt{1/2}|\langle{\eta^s|A_{\pi^-}|\hat \delta^{s+}}\rangle|$
$\sim 0.6$  
in {\bf II}. Using the above results on the asymptotic matrix 
elements, the value of $\epsilon$ in Eq.~(\ref{eq:epsilon}) and 
$\theta_P=-20^\circ$, we have the rates for the isospin non-conserving 
decays, 
$\Gamma(D_{s0}^{*+}\rightarrow D_s^+\pi^0)_{SU_f(4)}
 \simeq \Gamma(\hat F_0^+\rightarrow D_s^+\pi^0)_{SU_f(4)} 
\simeq 0.6\,\,  {\rm keV}$.  
These results are much smaller than the rates for the radiative decays 
of the charm-strange scalar mesons listed in Table~I, as conjectured 
in {\bf I}. Eventually, the ratios of decay rates under consideration 
can be obtained as 
(i) $R(\hat D_{s0}^{*+})\simeq 60$, (ii) $R(\hat F_0^+)\simeq 7$ and  
(iii) $R(\hat F_I^+)\simeq 0.005$ in Eq.~(\ref{eq:ratio-F_I}). 
In this way, it is seen that the experimental constraint 
Eq.~(\ref{eq:constraint-D_{s0}}) can be satisfied only in the case 
(iii). (For more details, see Refs.~\cite{Trento,HT-isospin}) Its 
assignment to an iso-singlet $DK$ molecule~\cite{BCL} has already been 
rejected~\cite{MS} because it leads to 
$R(\{DK\}) \gg R(D_{s0}^+(2317))_{\rm exp}$ 
as in (ii). Thus we conclude that assigning $D_{s0}^+(2317)$ to 
$\hat F_I^+$ is favored by the experiments while its assignment to the 
$I=0$ state, the conventional scalar $D_{s0}^{*+}$ or the scalar 
four-quark $\hat F_0^+$ (or the $DK$ molecule), is not favored. 

\section{Production of Charm-strange Scalar Mesons} 

Although assigning $D_{s0}^+(2317)$ to $\hat F_I^+$ is favored by 
experiments as seen above, its neutral and doubly charged partners, 
$\hat F_I^0$ and $\hat F_I^{++}$, have not yet been observed by 
inclusive $e^+e^-$ annihilation experiment~\cite{BABAR-search}. 
Therefore, we now study productions of charm-strange scalar four-quark 
mesons ($\hat F_I^{++,+,0}$ and $\hat F_0^+$).  To this aim, we consider 
their production through weak interactions, as a possible candidate, 
because OZI-rule violating creations of multiple $q\bar q$-pairs and 
their recombinations into four-quark meson states are expected to be 
strongly suppressed at high energies~\cite{Terasaki-production}. We, 
first, recall the so-called BSW Hamiltonian~\cite{BSW} as the effective 
weak Hamiltonian, 
\begin{equation}
H_w^{\rm BSW} \propto {a_1Q_1 + a_2Q_2 + \cdots + H_w' + h.c.,} 
                                                       \label{eq:BSW}
\end{equation}
where $Q_1$ and $Q_2$ are four-quark operators given by products of 
neutral and charged currents, respectively, and provide amplitudes for 
color suppressed and color favored decays, respectively, under the 
factorization prescription. The extra term $H_w'$ is automatically 
induced when the BSW Hamiltonian is obtained. It is given by a sum of 
products of colored currents and provides a non-factorizable amplitude, 
so that it is usually taken away. However, in this talk, it is left 
intact~\cite{Terasaki-hybrid-B,Terasaki-hybrid} because it can play an 
important role in production of charm-strange scalar four-quark mesons. 

Next, we draw quark-line diagrams within the minimal $q\bar q$-pair 
creation, because multiple $q\bar q$-pair creation is expected to be 
suppressed due to the OZI rule. In this approximation, the quark-line 
diagrams related to production of charm-strange scalar four-quark mesons 
in $e^+e^-\rightarrow c\bar c$ annihilation are given in Fig.~1. 
\begin{figure}[!t]  %
\begin{center} \hspace{5mm}
\includegraphics[width=135mm,clip]{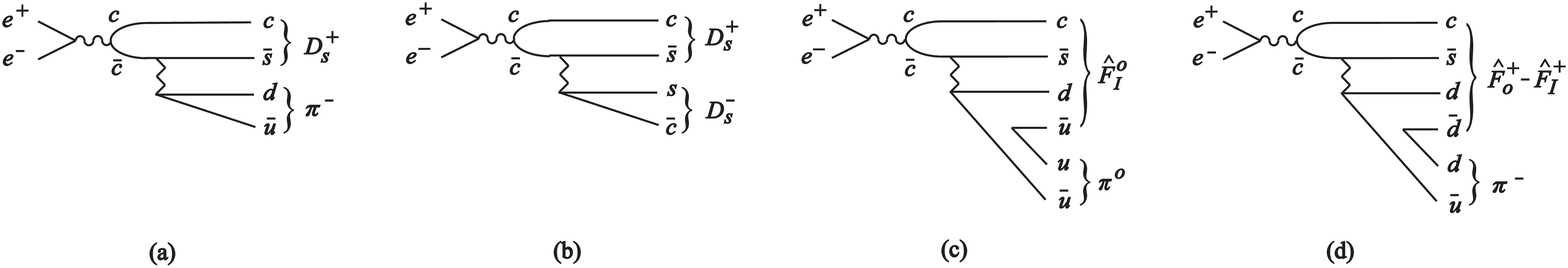}
\label{fig:vir-c-cbar.eps}       %
\end{center} 
\begin{quote}  
Fig.~1. 
Productions of charm-strange scalar mesons through 
$e^+e^-\rightarrow c\bar c$ within the minimal $q\bar q$-pair 
creation. (a) and (b) describe productions of $D_s^+\pi^-$, 
$D_s^{*+}\pi^-$, $D_s^+\rho^-$, etc. and $D_s^+D_s^-$, $D_s^{*+}D_s^-$, 
$D_s^+D_s^{*-}$, etc., respectively. Productions of $\hat F_I^0\pi^0$ 
and $(\hat F_0^+,\,\hat F_I^+)\pi^-$ are given by (c) and (d), 
respectively.  
\end{quote}
\end{figure}%
\begin{figure}[!b]     %
\begin{center}  
\includegraphics[width=135mm,clip]{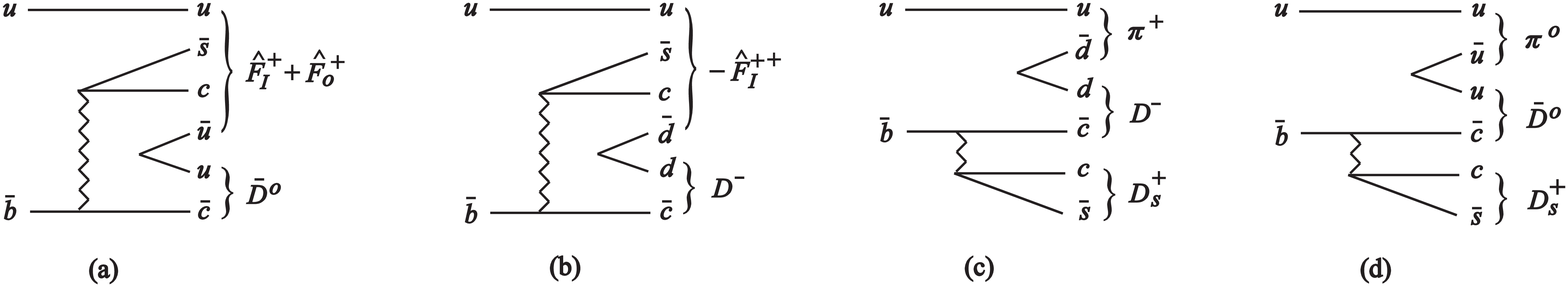}
\label{fig:via-Bu.eps}              %
\end{center} 
\begin{quote}  
{   
Fig.~2. 
Productions of charm-strange scalar mesons in weak decays of 
$B_u$ meson. (a) describes a production of $\hat F_I^+$ and 
$\hat F_0^+$ with $\bar D^0\,({\rm or}\,\bar D^{*0})$, 
(b) a production of $\hat F_I^{++}$ with $D^-\,({\rm or}\,D^{*-})$, and  
(c) and (d) productions of $D_{s}^+\pi^+$ with $D^-$ and 
$D_{s}^+\pi^0$ with $\bar D^0$, respectively.
}
\end{quote}
\end{figure}%
Because there is no diagram to describe production of $\hat F_I^{++}$  
in this approximation, as seen in Fig.~1, it is understood why the 
$e^+e^- \rightarrow c\bar c$ experiment~\cite{BABAR-search} found no 
evidence for it. Productions of $\hat F_I^0$, $\hat F_I^+$ and 
$\hat F_0^+$ mesons are described by Figs.~1(c) and (d). The diagrams 
Figs.~1(a) and (b) in which the weak vertices are given by the color 
favored spectator diagrams describe productions of $D_s^+\pi^-$, 
$D_s^{*+}\pi^-$, $D_s^+\rho^-$, etc. and $D_s^+D_s^-$, $D_s^+D_s^{*-}$, 
$D_s^{*+}D_s^-$, etc., respectively. By the way, it is known that color 
favored spectator decays are much stronger than color mismatched decays 
under the factorization prescription (i.e., 
$|a_1/a_2|^2 \simeq 6.8\times 10^{-3}$ at the scale of charm 
mass~\cite{Neubert}). In addition, 
non-factorizable contributions are actually small in hadronic weak 
decays of $B$ mesons~\cite{Terasaki-hybrid-B}, and they will be much 
smaller at higher energies. As seen in Fig.~1, productions of 
$\hat F_I^{+,0}$ and $\hat F_0^+$ involve rearrangements of colors and 
their amplitudes are non-factorizable, so that they will be much more 
strongly suppressed than the color favored processes. Therefore, it is 
not very easy to extract the $\hat F_I^0 \rightarrow D_s^+\pi^-$ signals 
in {\it inclusive} $e^+e^-\rightarrow c\bar c$ experiments. In the case 
of $\hat F_I^+$, however, one does not need to worry about large numbers 
of background events from Figs.~1(a) and (b) because its main decay is 
$\hat F_I^+\rightarrow D_s^+\pi^0$. Nevertheless, its evidence has not 
been observed in the radiative channel, because its decay into 
$D_s^{*+}\gamma$ is strongly suppressed as seen in {\bf III}. As for 
$\hat F_0^+$, it can decay much more strongly into $D_s^{*+}\gamma$ than 
$D_s^+\pi^0$ as seen in {\bf III}, although its production is depicted 
by the same diagram Fig.~1(d) as the production of $\hat F_I^+$. 
Therefore, reconstruction of $\hat F_0^+\rightarrow D_s^{*+}\gamma$ 
might be suspected to be efficient to search for $\hat F_0^+$. However, 
very large numbers of $D_s^{*+}$ and $\gamma$ (from 
$D_s^{*-} \rightarrow D_s^-\gamma$) produced through the spectator 
diagrams Figs.~1(a) and (b) (and in 
$e^+e^-\rightarrow c\bar c\rightarrow D_s^{(*)+}D_s^{(*)-}$, etc.   
without weak interactions) obscure the above signal $D_s^{*+}\gamma$. 
In this way, it will be understood that whether each of charm-strange 
scalar mesons can be observed or not depends on its production 
mechanism, and, therefore, it seems that no evidence for $\hat F_I^0$ 
and $\hat F_I^{++}$ in inclusive $e^+e^-\rightarrow c\bar c$ 
annihilation experiments does not necessarily imply their non-existence. 

Because it is difficult to observe $\hat F_I^{++}$,  $\hat F_I^{0}$ 
and $\hat F_0^{+}$ in inclusive $e^+e^-\rightarrow c\bar c$ experiments 
as seen above, we now study productions of charm-strange scalar 
four-quark mesons in $B$ decays. For this purpose, we again draw 
quark-line diagrams describing their productions within the minimal 
$q\bar q$-pair creation. As expected in the quark-line diagrams of 
Figs.~2 and 3, resonance peaks which are approximately degenerate with 
$D_{s0}^+(2317)$ have been observed in the following hadronic weak 
decays of $B$ mesons: 
$B_u^+ \rightarrow \bar D^0\tilde D_{s0}^+(2317)
[D_s^+\pi^0, D_s^{*+}\gamma]$ 
and 
$B_d^0 \rightarrow D^-\tilde D_{s0}^+(2317)[D_s^+\pi^0, D_s^{*+}\gamma]$  
in the BELLE experiment~\cite{BELLE-D_{s0}}, and 
$B_u^+\rightarrow 
\bar D^0({\rm or}\,\bar D^{*0})\tilde D_{s0}^+(2317)[D_s^+\pi^0]$  
and 
$B_d^0\rightarrow D^-({\rm or}\,D^{*-})\tilde 
D_{s0}^+(2317)[D_s^+\pi^0]$ 
in the BABAR experiment~\cite{BABAR-D_{s0}-B}. It should be noted 
that indications of new resonances have been observed in the 
$D_s^{*+}\gamma$ channel. It is quite different from the case 
of inclusive $e^+e^-\rightarrow c\bar c$. Therefore, the new resonances 
have been denoted by $\tilde D_{s0}^+(2317)$[observed channel(s)] to distinguish them from the previous $D_{s0}^+(2317)$. Because Figs.~2(a) 
and 3(b) involve both $\hat F_I^+$ and $\hat F_0^+$ and their main 
decays are quite different from each other, the new resonance can be assigned to $\hat F_I^+$ when it is observed in the $D_s^+\pi^0$ 
channel, while it might be assigned to $\hat F_0^+$ when it is observed 
in the $D_s^{*+}\gamma$ channel. Observations of $\hat F_I^{++}$ and 
$\hat F_I^0$ are expected in the process 
$B_u^+\rightarrow D^-({\rm or}\,D^{*-})\hat F_I^{++}[D_s^+\pi^+]$ 
as depicted in Fig.~2(b), and in the process 
$B_d^0\rightarrow \bar D^0\hat F_I^0[D_s^+\pi^-]$ 
as depicted in Fig.~3(a), respectively. Because the diagrams Figs.~2(a), 
2(b), 3(a) and 3(b) are of the same type, rates for production of 
$\hat F_I^{++}$ and $\hat F_I^0$ are expected to be not very far from 
that for $\tilde D_{s0}^+(2317)[D_s^+\pi^0]$, i.e., 
\begin{eqnarray}
&&
{B}(B_u^+\rightarrow D^-({\rm or}\,D^{*-})\hat F_I^{++}[D_s^+\pi^+])
\sim {B}(B_d^0\rightarrow \bar D^0\,({\rm or}\,\bar D^{*0})
\hat F_I^0[D_s^+\pi^-])     \nonumber\\   
&&\hspace{20mm}
\sim {B}(B\rightarrow \bar D\,({\rm or}\,\bar D^{*})
\tilde D_{s0}^+(2317)[D_s^+\pi^0])_{\rm exp}    
\sim 10^{-3}.        
                                                  \label{eq:prod-rate}
\end{eqnarray}
Therefore, $\hat F_I^{++}$ and $\hat F_I^0$ could be observed in 
$B\rightarrow \bar D({\rm or}\,\bar D^{*})D_s^+\pi$ decays. 
\begin{figure}[!t]    %
\includegraphics[width=135mm,clip]{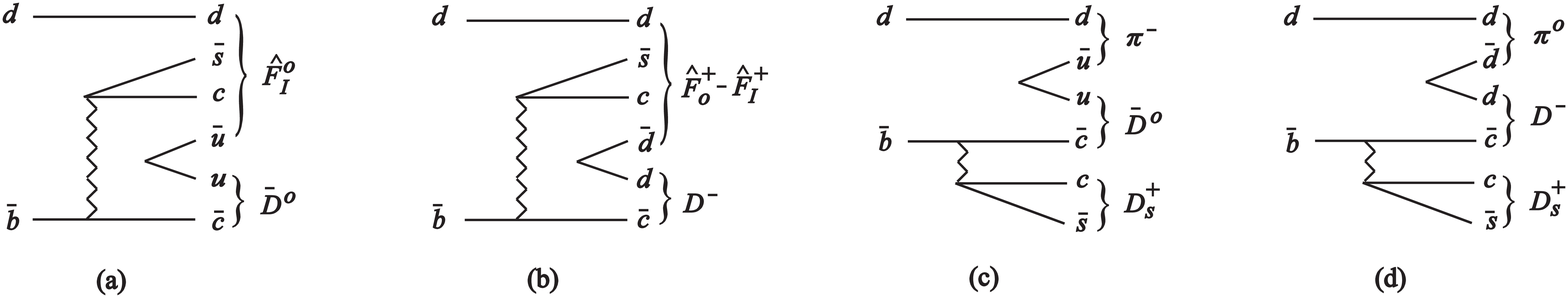}
\label{fig:via-Bd.eps}             %
\begin{quote} 
{   
Fig.~3. 
Productions of charm-strange scalar mesons in weak decays of 
$B_d$ meson. (a) describes a production of $\hat F_I^0$ with $\bar D^0$ 
(or $\bar D^{*0}$), (b) a production of $\hat F_I^+$ and $\hat F_0^+$ 
with $D^-$ (or $D^{*-}$). (c) and (d) provide productions of 
$D_s^+\pi^-$ with $\bar D^0$ and $D_s^+\pi^0$ with $D^-$, respectively.   
}
\end{quote}    
\end{figure}%

\section{Summary}

By studying the $D_{s0}^+(2317)\rightarrow D_s^+\pi^0$ and 
$D_{s0}^+(2317)\rightarrow D_s^{*+}\gamma$ decays, we have seen that 
assigning $D_{s0}^+(2317)$ to $\hat F_I^+$ is favored by experiments. 
To search for its partners $\hat F_I^0$ and $\hat F_I^{++}$, we have 
investigated productions of these four-quark mesons through hadronic 
weak interactions. As the results, we have found that detecting them 
in inclusive $e^+e^-\rightarrow c\bar c$ is likely quite difficult, 
although $D_{s0}^+(2317)$ itself has already been observed. 
Taking these points into consideration, we have estimated the branching 
fractions for decays of $B$ mesons producing $\hat F_I^{++}$ and  
$\hat F_I^0$ as 
${B}(B_u^+\rightarrow D^-\hat F_I^{++})
\sim {B}(B_d^0\rightarrow \bar D^0\hat F_I^{0}) \sim 10^{-3}$. 
As for observation of $\hat F_I^+$ and $\hat F_0^+$, we conclude that 
they could have been observed as resonances with approximately equal 
masses in two different channels, $D_s^+\pi^0$ and $D_s^{*+}\gamma$, 
as the BELLE collaboration observed. 

\section*{Acknowledgments} 
The author would like to thank the Yukawa Institute for Theoretical 
Physics at Kyoto University. Discussions during the YKIS2006 on 
"New Frontiers on QCD" were useful to complete this work. He also 
would like to appreciate the organizers for financial supports. 


\end{document}